# Single channel speech enhancement by colored spectrograms

Sania Gul[1,4], Muhammad Salman Khan[2#], Muhammad Fazeel[3]

[1]Department of Electrical Engineering, University of Engineering and Technology, Peshawar, Pakistan

[2]Department of Electrical Engineering, College of Engineering, Qatar University, Doha, Qatar

[3]Department of Mechanical Engineering, University of Engineering and Technology, Peshawar, Pakistan

[4]Artificial Intelligence in Healthcare, Intelligent Information Processing Lab, National Center of Artificial Intelligence, Peshawar, Pakistan

#Corresponding author: E-mail address: salman@qu.edu.qa

**Abstract:** *Speech enhancement concerns the processes required to remove unwanted background sounds from the target speech to improve its quality and intelligibility. In this paper, a novel approach for single-channel speech enhancement is presented, using colored spectrograms. We propose the use of a deep neural network (DNN) architecture adapted from the pix2pix generative adversarial network (GAN) and train it over colored spectrograms of speech to denoise them. After denoising, the colors of spectrograms are translated to magnitudes of short-time Fourier transform (STFT) using a shallow regression neural network. These estimated STFT magnitudes are later combined with the noisy phases to obtain an enhanced speech. The results show an improvement of almost 0.84 points in the perceptual evaluation of speech quality (PESQ) and 1% in the short-term objective intelligibility (STOI) over the unprocessed noisy data. The gain in quality and intelligibility over the unprocessed signal is almost equal to the gain achieved by the baseline methods used for comparison with the proposed model, but at a much reduced computational cost. The proposed solution offers a comparative PESQ score at almost 10 times reduced computational cost than a similar baseline model that has generated the highest PESQ score trained on grayscaled spectrograms, while it provides only a 1% deficit in STOI at 28 times reduced computational cost when compared to another baseline system based on convolutional neural network-GAN (CNN-GAN) that produces the most intelligible speech.*

A. Introduction

Speech enhancement (SE), a class of audio enhancement (AE) involves the techniques used for removing the unwanted sounds from the target speech, with minimum possible degradation on the target speech [1]. Speech enhancement is not only required for people with hearing problems, but may sometimes be required for normal listeners in applications like retrieval of vocals from the background music to be used for karaoke, retrieval of sound from events recorded in the noisy background e.g. on a seashore or a busy bus station, extracting the contents of a noisy voice-note, removing the high intensity echoes, or filling the gaps due to intrusion or damage usually encountered while recovering musical records from gramophone and tapes. With many emerging machine listening applications e.g. automatic speech recognition (ASR), automatic speaker identification (ASV), internet voice search engines, home assistants, and smart meeting rooms, the need to introduce more efficient SE techniques is increasing.

In recent years, the use of DNNs for SE has shown tremendous improvement over the classical methods [2]. Although there exist DNN models e.g. recurrent neural network (RNN), wave U-Nets, TasNet, and 1D convolutional neural networks (CNNs) that can process the audio directly without the need for its conversion to any other domain, it is found that converting the audio to a time-frequency (TF) domain representation, commonly called as a spectrogram, results in its better separation from the background noise as compared to processing it directly in time domain. This holds for human speech, animal sounds, music, and environmental sounds ([3], [4], [5], and [6]). There are a lot of image-processing DNNs that can provide AE by treating the spectrograms as images. These include 2D CNNs (e.g. in [7]), image U-Nets (e.g. [8], [9], [10], [11], and [12]), and image generative adversarial networks (GANs) (e.g. [13], [14], [1] and [15]). The advantage of using the spectrogram-based DNN models includes a lesser number of trainable network parameters, and a lesser training cost than the waveform-based models [16].

Image GAN [17] (we use the term 'image' to differentiate these GANs from the audio GANs that accept the audio directly in the time domain e.g. [18]) is a generative modeling technique initially applied to computer vision [14]. GAN tries to learn the probability density function (PDF) from the given training data and then generates new samples according to that PDF [19]. It generates new data samples similar to the real data by setting two neural networks (the generator and the discriminator) against each other. The generator tries to deceive the discriminator by generating fake samples according to the PDF of the real data. The discriminator, on the other hand, is usually a binary classifier that tries to differentiate between real and generated samples as accurately as possible [19]. Recent research has shown the potential of exploiting GANs in SE-related applications where they learn a suitable mapping function and accurately reconstruct the enhanced speech while maintaining speech quality and intelligibility [14].

*A. Recent Work*

The use of image GANs for audio was pioneered by [13], where pix2pix conditional GAN (cGAN) [20] is used for SE required for the ASV application. Pix2pix is a framework proposed originally for image-to-image translation [20], but it is later used for a wide range of audio applications including music generation [21] and [22], music inpainting [23], music transcription [24], sound anomaly detection [25], SE [14], ASR [26], voice conversion [27], speech inpainting [28] and sound source localization [29]. The models utilizing pix2pix for SE e.g. [13], and [14] use a complete pix2pix network i.e. both generator and discriminator while the SE model proposed in [1] uses DNN architecture resembling closely the generator side of the pix2pix cGAN. The generator of pix2pix has a U-Net [30] architecture, which accepts a noisy spectrogram at its input and maps it to a clean spectrogram. In our proposed SE model, we will also use only the generator side of the pix2pix network for spectrogram denoising. In all AE applications using the image GAN, the audio is required to be converted to a spectrogram before being processed by GANs. The spectrogram can be of any type (simple or log-scaled short-time Fourier transform (STFT), gammatonegrams, Mel spectrograms, scalograms, or others) according to the application's need. In all these cases the spectrograms are matrices of numbers (complex or real), which are treated as a grayscale image by the network. To the best of our knowledge, no AE application (e.g. source separation, inpainting, denoising, and dereverberation) has ever used colored spectrograms. In this paper, the colored spectrograms are investigated for the first time for speech denoising. It will be shown that using the color information provides appreciable improvement in speech quality and intelligibility at a much reduced computational cost than required by using the grayscaled images (TF matrices).

The motivation behind the use of colors for SE is based on the observation of their valuable effect in different sound classification tasks e.g. in pathology detection [31] and emotion recognition [32]. In [31], using the colored spectrograms has been found beneficial for the classification of different lung diseases from their scalograms (a type of spectrogram for slowly varying signals e.g. auscultation sounds) [31]. In automatic speech emotion recognition (SER) model [32], different colors show their strength in improving the accuracy of recognition for detecting different emotions from sounds e.g. joy and anger are better classified by the red color, while boredom by the green color.

### B. *Our contribution*

In this paper, we use the colored spectrogram for the first time for the speech denoising task and in fact for the first time for any kind of AE application. Looking at the spectrograms of noisy and clean speeches, some colormaps are found to be more effective than others in discriminating the noise contents from the speech contents. For example, as shown in Figure 1, the 'parula' map effectively displays the energy contents of speech to the human eye, 'pink' performs moderately, while the 'prism' completely fails to do so.

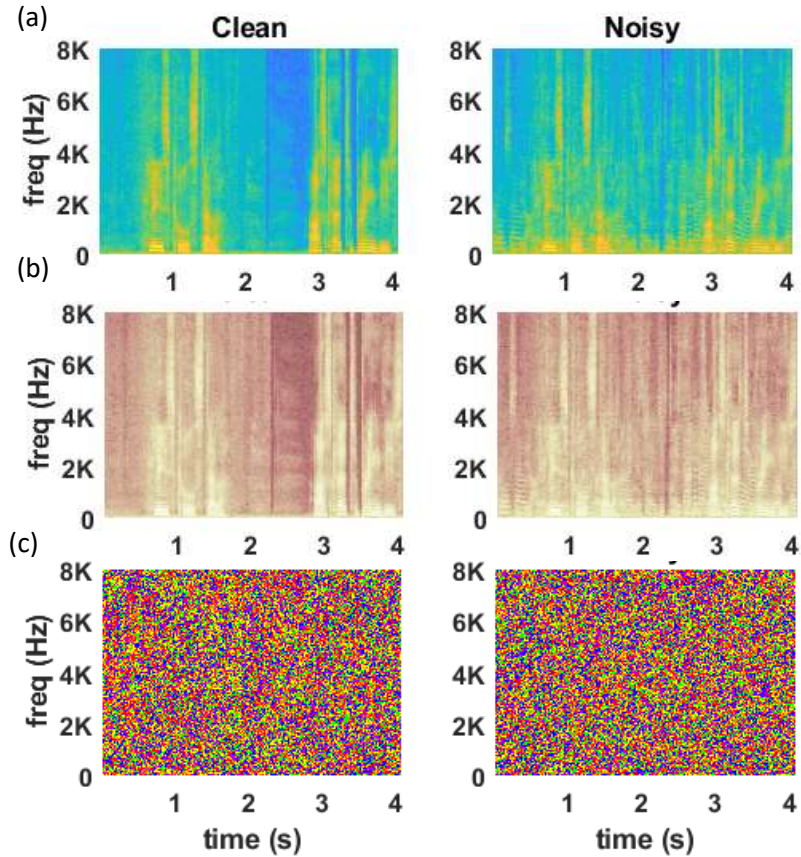

**Figure 1:** Spectrograms using different colormaps. **(a).** Parula, **(b).** Pink, **(c).** Prism

In this paper, the effect of colors on the 'computer eye' in automatically differentiating the pixels belonging to the speech or noise spectrum is explored. By 'computer eye' we mean the DNNs fundamentally developed for computer vision [33]. The colored spectrogram also shows the energy distribution across the spectrum at various time instants and hence the hue (darkness of color) of each pixel seems to be proportional to the magnitudes of STFT components. So following the human eye processing, we design a two-step model.

1. The first part takes a colored noisy spectrogram as input and denoises it by using a DNN having architecture that matches closely with the generator of the pix2pix GAN.

2. The second part converts the hue of each pixel to an STFT magnitude spectrogram by using regression neural network. The resulting spectrogram is then combined with the noisy phases and converted back to sound.

The rest of the paper is organized as follows. In the next section, an overview of the related work implementing speech denoising is given. The proposed system's overview is given in Section 3. In Section 4 the network settings, evaluation metrics, and the baseline algorithms for comparison with the proposed algorithm are described. Experimental results are presented in Section 5 and the paper is concluded in Section 6.

## 2. Related Work

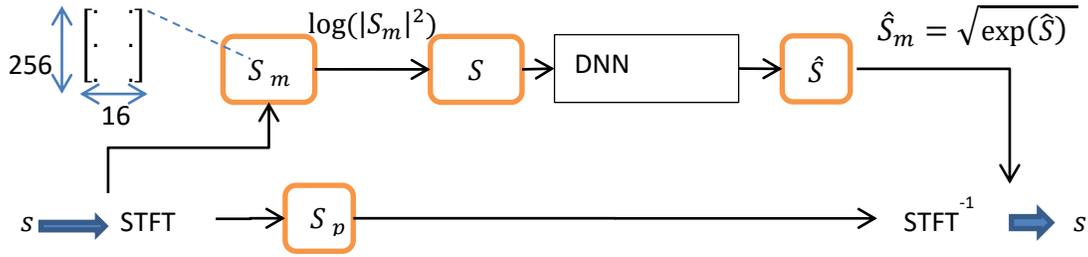

**Figure 2:** Block diagram of speech denoising model [1]

Our proposed model's design is based on the speech-denoising model of [1]. The block diagram of [1] is shown in Figure 2. In this model the time domain signal $s$ is converted to a 2D complex valued STFT spectrogram and the magnitude $S_m$ and phase $S_p$ are separately stored in two different matrices. To better discriminate the TF units with different energy levels and highlight the vanishing time-frequency units, the magnitude matrix $S_m$ is transformed to log powered spectrogram (LPS) $S$, by applying equation (1):

$$S = \log(|S_m|^2) \qquad (1), \qquad where \log(.) \; is \; element-wise \; natural \; log.$$

A DNN having an architecture similar to the pix2pix generator is then trained on a large number of noisy and clean LPS spectrograms, which would (after training) generate a clean LPS spectrogram, when a noisy spectrogram is given at its input. The generated LPS spectrogram $\hat{S}$ is later transformed back to $\hat{S}_m$ by using equation (2) and then combined with the noisy phase matrix $S_p$ to obtain the STFT matrix, which is then converted to estimated speech $\hat{s}$, by taking its inverse STFT (ISTFT).

$$\hat{S}_m = \sqrt{\exp(\hat{S})} \qquad (2)$$

## 3. System overview

### A. Proposed Model

The proposed model is named as 'C-Net', where 'C' represents 'color' and 'Net' the DNN. The signal-flow model of C-Net is shown in Figure 3.

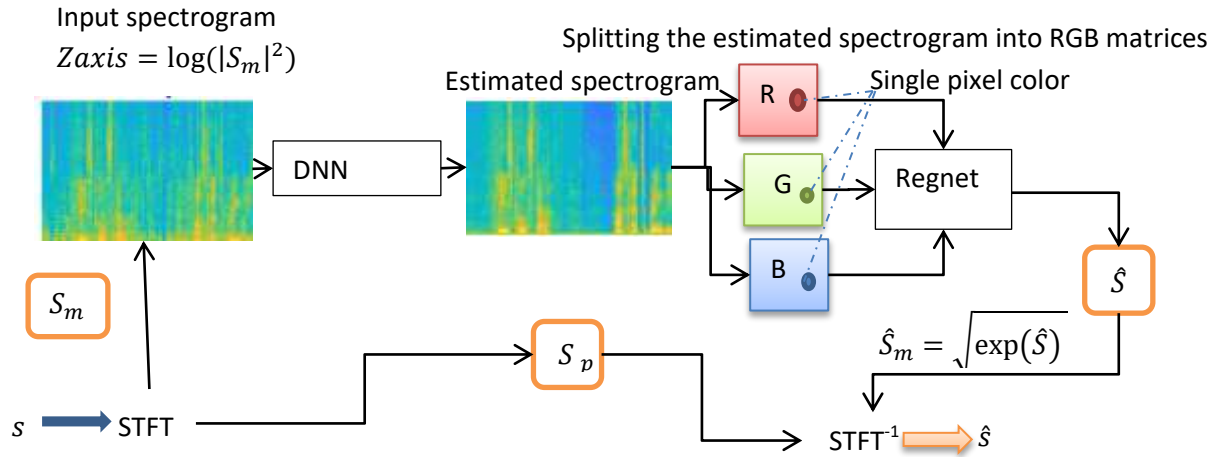

**Figure 3:** Signal-flow diagram of C-Net

The noisy time domain signal $s$ is converted into a colored spectrogram and given as an input to the DNN. The DNN is trained on thousands of such colored spectrograms of noisy and clean speech. The clean spectrogram acts as ground truth (GT) for network training. During the testing phase, the DNN regenerates an estimated clean spectrogram, when a noisy spectrogram is given at its input. This estimated clean spectrogram is a colored image, where each pixel is composed of three colors 1) red (R), 2) green (G), and 3) blue (B). These colors act as input to a multiple linear regression neural network (regnet), trained to predict the log-squared magnitudes corresponding to each pixel's color contents. The process of predicting the log-squared magnitudes from colors is done individually for all the pixels and the resulting outputs for all the pixels (belonging to the same input image) are again arranged in a matrix form. As STFT is symmetric around direct current (DC; the zero frequency component of signal), the matrix is flipped around DC to obtain the missing half of the LPS matrix $\hat{S}$ and equation (2) is used to obtain the estimated STFT magnitude matrix $\hat{S}_m$. This matrix is then combined with the noisy phase matrix $S_p$, and converted from polar to Cartesian form. Then ISTFT is applied over it to obtain the estimated speech $\hat{s}$.

B.  *Deep neural network (DNN) architecture*

The DNN architecture of C-Net is shown in Figure 4.

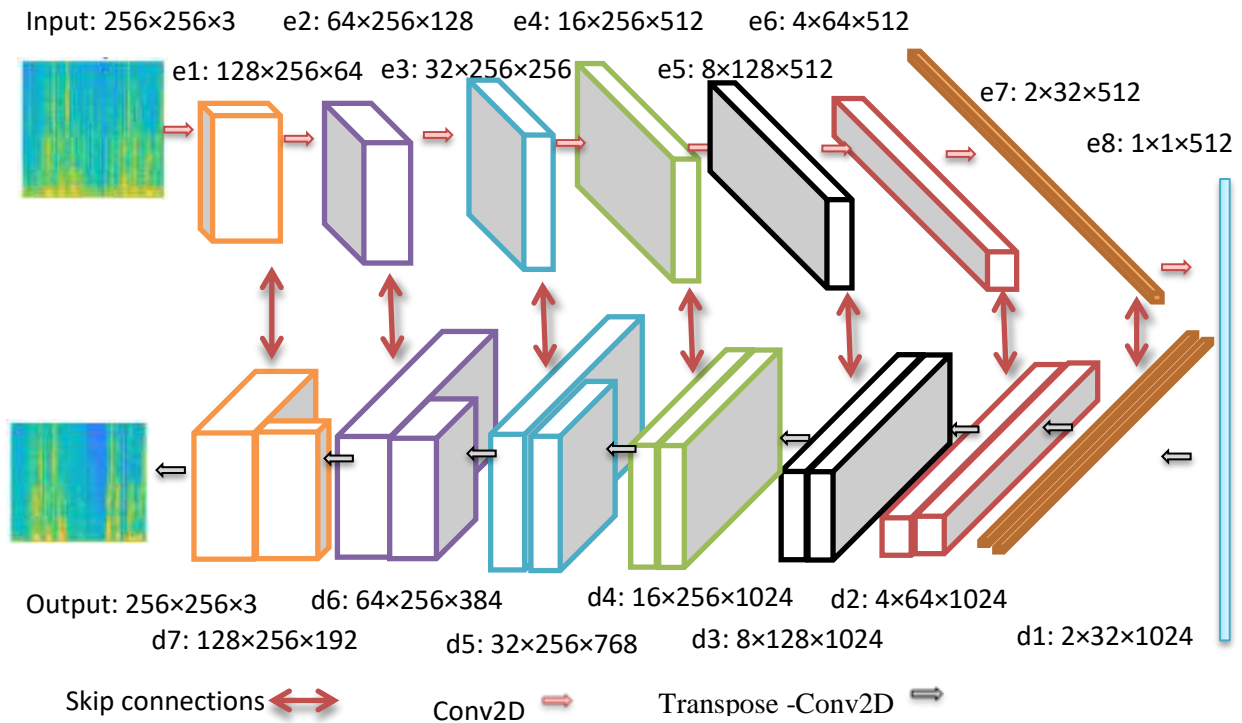

**Figure 4:** The architecture of C-Net

The encoder of the DNN designed for C-Net has the same architecture as in [1], originally adapted from the generator model of pix2pix GAN. The width of the spectrogram, given as an input, corresponds to time-domain signal's duration, its height represents the frequencies present in the audio signal while the number of channels shows whether it is a grayscaled (channel = 1) or a colored (channel = 3) image. The model of [1] has replaced the transposed-convolution layers at the decoder side with sub-pixel convolution layers introduced in [34] and changed the input size from 256×256×3 (width × height × channels) to 16×256×1 to achieve low-latency required for real-time applications. However, these modifications were not made in C-Net and the original parameters of the pix2pix generator model are retained, as the proposed model is not intended to tackle the real-time applications.

Both the encoder and decoder sides consist of eight (8) layers each. Each encoder layer consists of a conv2D layer, followed by batch normalization and a leaky RELU layer. The number of feature channels is 64, 128, and 256 for e1 to e3 respectively, while it is 512 for the rest of the encoder layers. The stride and kernel values for each encoder layer are kept same as in [1], i.e. (1, 2) for the first 4 layers (e1 – e4) and (2, 2) for the last 4 layers (e5 –e8). The filter size is kept as (5, 7) for e1-e3, (5, 5) for e4 and e5, and (3, 3) for the last 2 layers. It is shown in [1] and [12] that the use of asymmetric filters performs better than the symmetric filters, to capture the important spectral patterns of speech. All stride, filters, and feature channels are symmetric at the corresponding decoder layers connected by skip connections to the encoder side. Also, a drop-out of 50% is applied to the first three layers of the decoder (d1 –d3) to avoid overfitting.

Log-spectral distance (LSD) is used for training, as it gives slightly better results than the conventional L1 and L2 losses [1]. The LSD loss function is given by equation (3) as:

$$loss_{LSD} = \frac{1}{T}\sum_{i=1}^{T}\sqrt{\frac{1}{F}\sum_{j=1}^{F}[S(i,j) - \hat{S}(i,j)]^2} \tag{3}$$

Where $T$ represents the total time frames and $F$ represents total frequency bands, and $i$ and $j$ show time and frequency indices for each frame in a spectrogram. The network trainable parameters are 64,975,171, which exceeds by almost 6000 than those required by an equivalent network, working on grayscaled images.

### C. Regression neural network architecture

The regression neural network (regnet) is a fully connected (FC) feed-forward shallow NN. It has three inputs, for reading the hue (RGB contents) of each pixel of the estimated spectrogram (produced at the output of DNN) and translates them into the corresponding log-squared magnitudes, which are then again arranged into an LPS spectrogram.

The regnet has three hidden layers (HLs), each having 10 neurons. Each HL has a 'tanh' activation function, while the transfer function of the output layer is 'linear'. The input predictor variables R, G, and B are standardized i.e. they are centered and scaled by the corresponding column mean and standard deviation. The regnet has a total of 240 learning parameters, which contributes minutely to the total learning parameters of C-Net.

## 4. Network parameters

### A. Dataset

The dataset used for the training and testing of C-Net is publicly available [35]. It consists of clean and noisy data, sampled at 48 kHz. There are 2 datasets available in this corpus [35], 1) the '28 speaker' and 2) the '56 speaker' set. C-Net uses the '28 speaker' set. In this set, utterances from 28 speakers are reserved for training, while 2 are kept for testing. This dataset is created by selecting the clean utterances of 30 native English speakers (gender-wise balanced) from the voicebank corpus [36]. Each speaker has uttered 400 sentences. The noisy dataset is formed by adding 10 different types of noises (8 real and 2 synthetic) taken from the Demand dataset [37]. Four signal-to-noise ratio (SNR) values are used, which are 15, 10, 5, and 0dB. This results in 40 (10 types of noise × 4 SNRs) different types of conditions for the training set. The duration of the training set is around 9.5 hours.

For testing, 5 types of noises are selected from [37] and mixed with the clean speech at SNR values of 17.5, 12.5, 7.5, and 2.5dB. This results in 20 different conditions for the test set. The duration of the testing set is around 30 minutes. All noise types and conditions are mutually exclusive between the training and the testing datasets. Both datasets are down-sampled to 16 KHz and clean and noisy data from all speakers belonging to the same dataset are concatenated in two separate audio files, before being transformed into spectrograms.

### B. Spectrogram creation

The colored spectrograms required for the training and testing the C-Net are formed by taking the STFT of the input signal, with the parameters listed in Table 1.

**Table 1:** Spectrogram parameters

| STFT parameters | Values |
|---|---|
| Window Shape | Hanning |
| STFT frame length | 512 samples (32ms) |
| Hop size | 256 samples (16ms) |

The time domain signal $s$ is converted into a colored spectrogram with the x-axis representing time, the y-axis frequency, and z-axis the square of magnitude on the logarithmic scale, as given by equation (1) and shown in Figure 5(a).

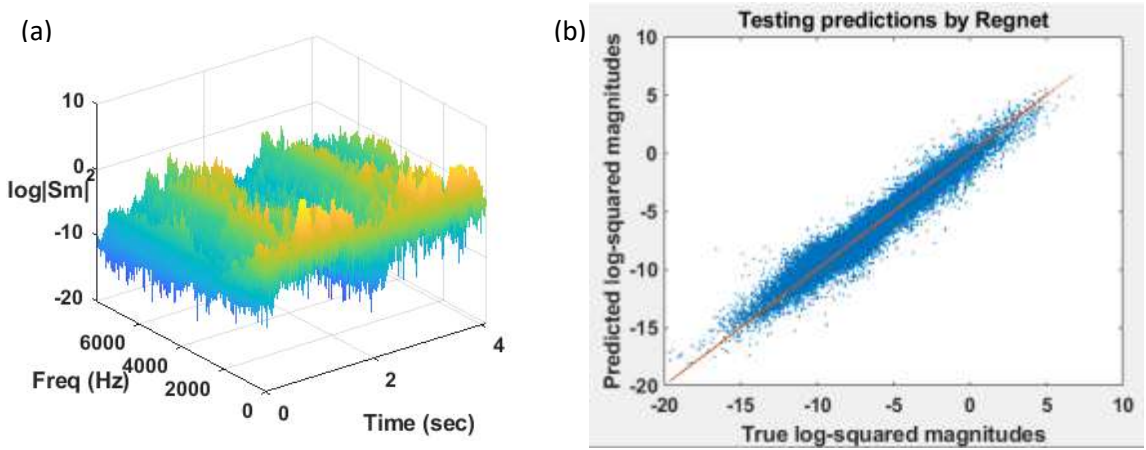

**Figure 5**: **(a)** Spectrogram axis description. **(b).** Fitting accuracy of regression network

Including DC, only 257-point STFT magnitudes are plotted for positive frequencies in the spectrogram. As the other half of STFT for the negative frequencies is symmetric around DC, it is neither plotted nor required for training. The information in this half is redundant; also its inclusion would increase the image size and the learning parameters of DNN, which would increase the training time and complexity.

As the input image required by the DNN must have the dimensions in the power of 2, the last (257[th]) row of the spectrogram is cropped, which covers the highest 31.25Hz band and has a little role in speech quality [1]. After the testing phase, when the full STFT matrix is reconstructed at the final stage, this highest frequency, removed at the input of DNN, would be required. It is found that instead of using a row of all zeros, the duplication of values from the 256[th] row would generate slightly better results. The missing half of the STFT matrix for negative frequencies is calculated by flipping the STFT magnitude matrix around DC, as described above.

At the sampling frequency of 16 KHz, a clip of 4.12 seconds is required to make the width of the spectrogram equal to 256. As C-Net uses a colored spectrogram, the image size at the input of DNN is 256×256×3. For spectrogram storage on a hard disk, the method described in [38] is used. However, the image's dimensions are changed after storage. This shortcoming was managed by first storing the images (one at a time) in the current directory using [38], retrieving them from there, resizing them to a size of 256×256×3 and then storing them again as jpg images in separate training and testing folders (according to the dataset from which they were created), using the '*imwrite*' instruction. These colored spectrograms are normalized to have zero mean and unit variance, before being given to the DNN.

C. *Training and Testing parameters of DNN*

The training and testing parameters of DNN are listed in Table 2.

**Table 2:** DNN parameters

| DNN parameters | Values |
| --- | --- |
| Training samples (noisy/clean) | 8205 each |
| Testing samples (noisy/clean) | 502 each |
| Input dimensions (width × height × channels) | 256×256×3 |
| Weight initialization | $N(\mu, \sigma) = N(0, 0.02)$ |
| Optimizer | Adam |
| Decay rates | $\beta_1 = 0.5, \beta_2 = 0.9$ |
| Training Steps | 6000 |
| Initial learning rate | 0.0001 |
| Batch size | 1 |
| Loss function | LSD (eq. (3)) |

Instead of epochs, the training loop is setup to work in 'steps', as in the original pix2pix model [20].

D. *Training parameters of regnet*

For regnet training, the RGB values of all pixels in a 256 ×256 colored spectrogram are stored in three columns of an Excel sheet, while its fourth column contains the log-squared magnitude of the signal corresponding to each pixel.

From one image, 65,536 samples are generated, as the image size is 256 ×256. The training samples are collected from 10 such clean spectrograms of the training dataset, resulting in a total of (10×65536) 0.6 million training samples. For regnet, a hold-out validation method is used, where 80% of the dataset is used for training and the remaining 20% is used for testing. The training is done for 1000 epochs. The testing mean square error (MSE) was 0.15 as shown in Figure 5(b). After training and testing on clean spectrograms, regnet is inserted in the C-Net model of Figure 3, where it predicts the log-squared magnitudes, corresponding to the RGB contents of each pixel of the estimated spectrogram produced by the generator.

*E. Evaluation metrics*

The metrics used for the evaluation of quality and intelligibility of the estimated speech $\hat{s}$ at the output of C-Net are the perceptual evaluation of speech quality (PESQ) [39] and short-term objective intelligibility (STOI) [40] respectively. These metrics are chosen as they are used by the baseline algorithms (described in the next section) in this paper. The PESQ score lies between -0.5 to 4.5, while the STOI lies between 0 and 1, usually expressed in %.

*F. Baseline algorithm*

Four state-of-the-art baseline methods are used for the comparison of the proposed algorithm. They are selected as they are trained and tested on the same dataset, as was used by the C-Net. The results of these models are reported directly from their papers.

The first baseline algorithm, used for the comparison of C-Net, is a single-channel speech denoising model proposed in [1]. This model is chosen as its architecture is very much similar to C-Net. Its working is already explained in section 2. The input must be in the form of LPS [1]. The second algorithm chosen for comparison is the speech enhancement generative adversarial network (SEGAN) [18], which processes the signal directly in the time domain, using the generative adversarial network (GAN). SEGAN is a 1D adaptation of the pix2pix model operating on a time-domain speech signal [30]. The third speech-denoising model [15] uses a spiking neural network in U-Net architecture. The individual neurons in this network emit a spike when their membrane potential reaches the threshold value. It is a low-powered network useful for cell phones. The noisy input magnitude spectrogram is mapped to a cleaner version by the network and later the noisy phases are combined to produce enhanced speech. The model requires log-squared magnitude spectrograms at its input. The last model in the list of the baseline methods is [14], which uses a CNN-GAN, where the generator is a CNN-based autoencoder, while the discriminator is a binary classifier with architecture that is similar to the encoder side of the generator. The input must be in the form of a Gammatone spectrogram [14]. Except for SEGAN [18], which is a time domain algorithm, all other baseline algorithms and our proposed algorithm C-Net are: 1) TF-based and 2) use noisy phase for signal reconstruction.

## 5. Experiments and Results

*Case 1: Selection of optimal colormap*

To select the most optimum colormap for speech denoising from the 18 different options available in MATLAB 2019b [41], 18 sets of spectrograms are generated, each using a unique colormap, from the clean test speech signals. In an initial experiment of the most optimum colormap selection, the DNN is bypassed, as its training is very lengthy. Only the training of 19 separate regnets is carried out; each with two clean spectrograms taken from the training dataset of a particular colormap. The outputs obtained for each set are converted to speech signals and evaluated against the clean test speech. PESQ and STOI scores of these 18 sets are given in Table 3.

**Table 3:** Quality and intelligibility scores for each colormap

| SNo. | Colormap | PESQ | STOI |
|---|---|---|---|
| 1 | Parula | 1.6779 | 0.9459 |
| 2 | Autumn | 1.1992 | 0.7992 |
| 3 | Bone | 1.166 | 0.8032 |
| 4 | Color-cube | 1.049 | 0.6087 |
| 5 | Cool | 1.2561 | 0.8437 |
| 6 | Copper | 1.2001 | 0.8138 |
| 7 | Flag | 1.1603 | 0.3267 |
| 8 | Gray | 1.052 | 0.4832 |
| 9 | Hot | 1.2221 | 0.8501 |
| 10 | HSV | 1.21 | 0.8521 |
| 11 | Jet | 1.2074 | 0.8263 |
| 12 | Lines | 1.1539 | 0.4292 |
| 13 | Pink | 1.1539 | 0.4292 |
| 14 | Prism | 1.0624 | 0.2109 |
| 15 | Spring | 1.2038 | 0.8178 |
| 16 | Summer | 1.1871 | 0.7972 |
| 17 | Winter | 1.0694 | 0.7022 |
| 18 | Gray* | 1.0643 | 0.629 |

*In case of GRAY colormap, the regnet has only one input and one output.

As the output of 'parula' is better than others, it is selected for the final training and testing of C-Net.

*Case 2: Speech denoising by C-Net*

The duration of sound clips used for creating the spectrograms of C-Net is almost 16 times the duration used by [1]. As a result, the number of spectrograms, produced from the training and testing audios, is 16 times less than those present in the corresponding datasets [1]. C-Net is trained from 2000 to 8000 (in the increment of 1000) steps and evaluated after each increment. It was found that the best PESQ and STOI scores were obtained at 6000 steps. According to the parameters mentioned, the training samples of [1] would be 16 times greater than those of C-Net i.e. 8205 × 16 = 131280 training examples. Their batch size was 64, so in 1 epoch there would be approximately 131280/64 =2052 steps /iterations, and as their system is trained on 30 epochs, the step count reaches 2052 × 30 =61560, which is almost 10 times greater than those required by C-Net (6000 vs. 61560). In the case of [18], each training example is a chunk of 1.02 seconds, which results in the generation of 33400 training samples from the training dataset of around 9.5 hours duration. As their batch size was 400 and the number of training epochs was 86, it would result in (33400/400) × 86 = 7181 steps. So, its training cost is almost 1180 steps higher than the C-Net. Similarly, in [14], there are approximately 1.6 million training examples. So, in [14] the number of steps (according to their batch size of 200 and 20 epochs and window size of 20ms with 50% overlap) would be (1690230 /200) × 20 = 169023, which is almost 28 times greater than required for C-Net. Likewise, in [15], there are approximately 1 million samples for training (with a batch size of 32 and required epochs equal to 60), the total training steps required are (1056393/32) × 60 = 1,980,738 which are 330 times greater than those required for C-Net.

The results achieved by C-Net over the unprocessed data is given in Table 4.

**Table 4:** Results of C-Net

| Method | PESQ | STOI |
|---|---|---|
| Unprocessed | 1.40 | 0.89 |
| C-Net | 2.24 | 0.90 |

Although the dataset used by C-Net is the same as was used in the baseline models, there is a difference in the results reported for the unprocessed dataset. They have not mentioned whether they have applied evaluation metrics on the test audio signals after concatenating the audios together (as done for C-Net), by segmenting them in clips of a particular duration, or on per speaker basis. So, as the benchmark (results of unprocessed data) is different, instead of comparing the models directly, the gain achieved over the unprocessed data by each model, is compared in Table 5.

**Table 5:** Comparison of different models in terms of gain achieved over raw data

| Gain over unprocessed data | Δ PESQ | Δ STOI | Required training steps |
|---|---|---|---|
| C-Net | 0.84 | 0.01 | **6000** |
| [1] | **0.94** | 0.01 | 61560 |
| [18] | 0.26 | NA* | 7181 |
| [14] | 0.37 | **0.02** | 169,023 |
| [15] | 0.69 | 0 | 1,980,738 |

NA* = Not Available

The best results for each metric are boldfaced in Table 5. It can be seen from these results, that C-Net outperforms [18], [14], and [15] in terms of the quality gain over the unprocessed data by achieving 0.58, 0.47, and 0.15 points respectively. The STOI of [14] is highest, which is only 1% higher than C-Net. C-Net offers intelligibility equal to [1] and 1% higher than [15]. The computational cost of C-Net is the lowest among all the methods. C-Net requires almost 1200 steps less than those required by the SEGAN [18], and 10, 28, and 330 times less than are required for [1], [14], and [15] respectively without any noticeable effect in speech quality and intelligibility when compared to the top scorers ([1] for PESQ, and [14] for STOI) of the baseline methods. The possible cause of the insignificant drop in PESQ against [1] may be due to the differences in the clip duration of the two models. It is observed during the experiments that for shorter clip durations (as were used by [1]), the PESQ was higher. It is already mentioned in the earlier discussion that unlike [1], C-Net is not designed for low-latency applications, but it can be modified in the future to compensate such applications by adjusting the network parameters.

The time domain, spectrum and the spectrogram of noisy, clean and the estimated signals produced by C-Net, are shown in Figure 6 for comparison.

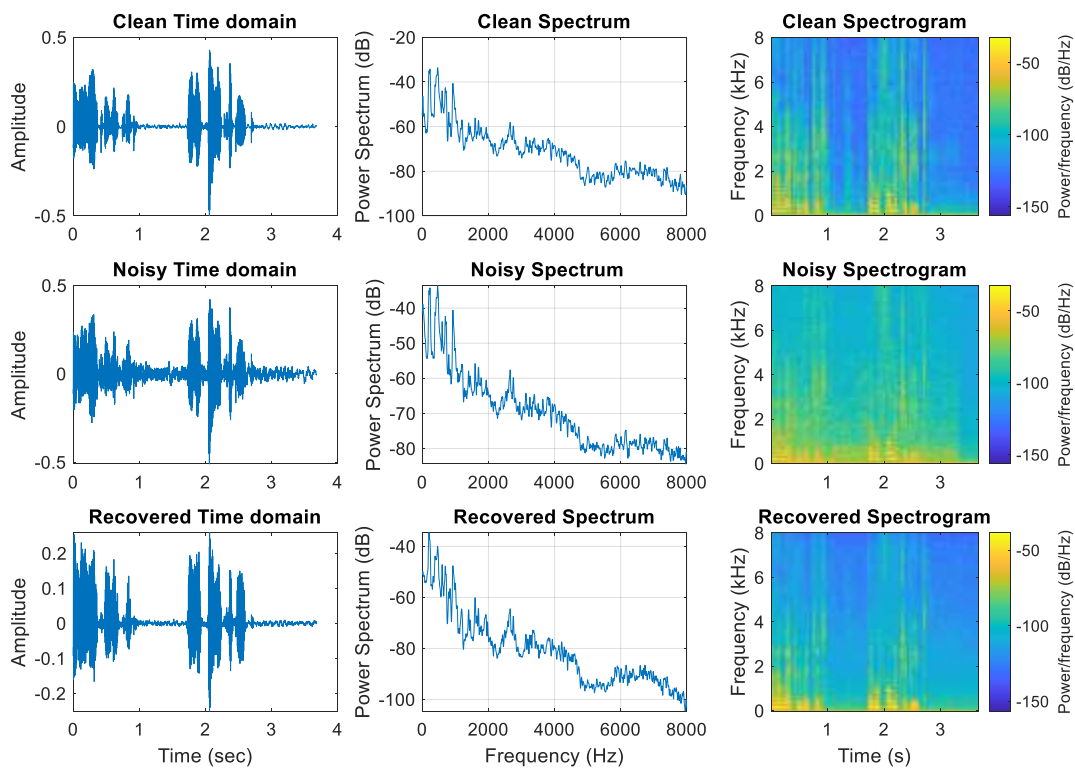

**Figure 6:** Clean, noisy and estimated (recovered) speech

### 6. Conclusion

In this paper, a novel idea of using colored spectrograms for speech-denoising applications is proposed. The color spectrogram is first denoised by using the pix2pix generator and later translated into an STFT magnitude matrix by the regression neural network. The gain in intelligibility over the unprocessed data is almost similar, at a much reduced computational cost than the baseline models. The gain in quality, over the raw signals, is also comparable to [1], the model producing the highest PESQ score. Although, in this paper, colors are explored for speech denoising, in the future, it would be interesting to investigate their use for other AE applications e.g. dereverberation, source separation, and source inpainting. Also, using more powerful image-denoising networks that estimate the phase along with magnitude instead of utilizing the noisy phase for signal reconstruction e.g. phase-aware SE models [42] and [43] may further elaborate the benefits of colors.

**References**

[1]. A. E. Bulut and K. Koishida, "Low-Latency Single Channel Speech Enhancement Using U-Net Convolutional Neural Networks," in Proc. *IEEE International Conference on Acoustics, Speech and Signal Processing (ICASSP)*, Barcelona, Spain, May 2020.

[2]. H. Purwins, B. Li, T. Virtanen, J. Schlüter, S. Chang and T. Sainath, "Deep Learning for Audio Signal Processing," in *IEEE Journal of Selected Topics in Signal Processing*, Vol. 13, no. 2, May 2019.

[3]. S. Rickard, "The DUET blind source separation algorithm," in *Blind Speech Separation, Signals and Communication Technology*, Netherlands: Springer, 2007.

[4]. Peter C Bermant, "BioCPPNet: Automatic Bioacoustic Source Separation with Deep Neural Networks," in *Scientific Reports*, Vol. 11(1), pp. 23502, Dec. 2021.

[5]. E. Cano, D. FitzGerald, A. Liutkus, M. D. Plumbley and F. Stöter, "Musical Source Separation: An Introduction," in *IEEE Signal Processing Magazine*, Vol. 36, no. 1, Jan. 2019.

[6]. S. A. Nossier, J. Wall, M. Moniri, C. Glackin and N. Cannings, "An Experimental Analysis of Deep Learning Architectures for Supervised Speech Enhancement," in *Electronics* 10, no. 1: 17, 2021.

[7]. S. Balasubramanian, R. Rajavel, Asuthos Kar, "Ideal ratio mask estimation based on cochleagram for audio-visual monaural speech enhancement," in Applied Acoustics,pp. 109524,Vol. 211,2023.

[8]. Sania Gul, M. S. Khan, and S. W. Shah, "Integration of deep learning with expectation maximization for spatial cue based speech separation in reverberant conditions," in *Applied Acoustic*s, Vol. 179, Aug. 2021.

[9]. Sania Gul, M. S. Khan, and S. W. Shah, "Preserving the beamforming effect for spatial cue-based pseudo-binaural dereverberation of a single source," in *Computer Speech & Language*, Vol. 77, pp. 101445, Jan. 2023.

[10]. Vinitha George, E., and V. P. Devassia, "A novel U-Net with dense block for drum signal separation from polyphonic music signal mixture," in *Signal, Image and Video Processing*, Vol. 17, no. 3, pp. 627-633, 2023.

[11]. G. W. Lee, K. M. Jeon and H. K. Kim, "U-Net-Based Single-Channel Wind Noise Reduction in Outdoor Environments," in Proc. *IEEE International Conference on Consumer Electronics (ICCE)*, Seoul, Korea, Jan. 2020.

[12]. O. Ernst, S. E. Chazan, S. Gannot and J. Goldberger, "Speech Dereverberation Using Fully Convolutional Networks," in Proc. *26th European Signal Processing Conference (EUSIPCO),* Rome, Italy, Sept. 2018.

[13]. Michelsanti, D., and Tan, Z, "Conditional Generative Adversarial Networks for Speech Enhancement and Noise-Robust Speaker Verification," in Proc. *Interspeech*, Stockholm, Sweden, Aug.2017.

[14]. N. Shah, H. A. Patil and M. H. Soni, "Time-Frequency Mask-based Speech Enhancement using Convolutional Generative Adversarial Network," in Proc. *Asia-Pacific Signal and Information Processing Association Annual Summit and Conference (APSIPA ASC),* pp. 1246-1251, Honolulu, HI, USA, 2018.

[15]. Riahi, Abir, and Éric Plourde, "Single Channel Speech Enhancement Using U-Net Spiking Neural Networks," in *arXiv preprint,* arXiv:2307.14464, 2023.

[16]. V. S. Kadandale, J. F. Montesinos, G. Haro and E. Gómez, "Multi-channel U-Net for Music Source Separation," in Proc. *IEEE 22nd International Workshop on Multimedia Signal Processing (MMSP)*, pp. 1-6, Tampere, Finland, 2020.

[17]. I. Goodfellow, J. Pouget-Abadie, M. Mirza, B. Xu, D. Warde-Farley, S. Ozair, et al., "Generative adversarial nets," in *Advances in Neural Information Processing Systems (NIPS),* Vol. 27, pp. 2672-2680, 2014.

[18]. Pascual, Santiago, Antonio Bonafonte, and Joan Serra. "SEGAN: Speech enhancement generative adversarial network," in *arXiv preprint arXiv:1703.09452* (2017).

[19]. Dash, Ankan, Junyi Ye, and Guiling Wang, "A review of Generative Adversarial Networks (GANs) and its applications in a wide variety of disciplines - From Medical to Remote Sensing," in *arXiv preprint,* arXiv:2110.01442, 2021.

[20]. P. Isola, J. Zhu, T. Zhou, and A. A. Efros, "Image-to-image translation with conditional adversarial networks," in Proc. *IEEE conference on computer vision and pattern recognition*, pp. 1125–1134, Honolulu, HI, USA, July. 2017. https://phillipi.github.io/pix2pix/

[21]. Siphocly NN, El-Horbaty ES, Salem AB, "Intelligent System Based on Deep Learning Technique for Accompaniment Music Generation," in Proc. The Institute of Electronics, Information and Communication Engineers (IEICE), Mar. 2021.

[22]. Rivas Ruzafa, Elena, "Pix2Pitch: generating music from paintings by using conditionals GANs," in *Ph.D. thesis*, Madrid, E.T.S. de Ingenieros Informáticos (UPM), (2020). [*Available online*]: https://oa.upm.es/view/institution/ETSI=5FInformatica/.

[23]. Sun-Kyung Lee, "Deep Generative Music Inpainting with Mel-Spectrogram ," in Ph.D. Thesis, KAIST, Daejeon, South Korea, 2020. [*Available online*]: https://mac.kaist.ac.kr/~juhan/gct634/2020-Fall/Finals/Deep_Generative_Music_Inpainting_with_Mel_Spectrogram.pdf

[24], Kim JW, Bello JP, "Adversarial learning for improved onsets and frames music transcription," in arXiv preprint, arXiv:1906.08512., 2019.

[25]. Kang C, Lee JH, Ji Y, Vu DM, Jung S, Kang C, "Real-world Application of pix2pix GAN-based Acoustic Signal Denoising for Enhanced Noise and Vibration Inspection in a Manufacturing Line," in research gate, Sept. 2023, doi:10.13140/RG.2.2.29122.35525.

[26]. C. Donahue, B. Li and R. Prabhavalkar, "Exploring Speech Enhancement with Generative Adversarial Networks for Robust Speech Recognition," in Proc. IEEE International Conference on Acoustics, Speech and Signal Processing (ICASSP), pp. 5024-5028, Calgary, AB, Canada, 2018.

[27]. Akadomari H, Sato Y, Kobayashi Y, "Comparison of the number of training data for Pix2Pix voice conversion system," in Proc. *IEEE 8th Global Conference on Consumer Electronics (GCCE)*, pp. 840-841, Oct. 2019.

[28]. Strods D, Smeaton AF, "Enhancing Gappy Speech Audio Signals with Generative Adversarial Networks," in *arXiv preprint,* arXiv:2305.05780. May 2023.

[29]. S. Kita and Y. Kajikawa, "Sound Source Localization Inside a Structure Under Semi-Supervised Conditions," in *IEEE/ACM Transactions on Audio, Speech, and Language Processing*, Vol. 31, pp. 1397-1408, 2023.

[30]. S. Abdulatif, K. Armanious, K. Guirguis, J. T. Sajeev and B. Yang, "AeGAN: Time-Frequency Speech Denoising via Generative Adversarial Networks," in Proc. *28th European Signal Processing Conference (EUSIPCO)*, pp. 451-455, Amsterdam, Netherlands, 2021.

[31]. S. B. Shuvo, S. N. Ali, S. I. Swapnil, T. Hasan and M. I. H. Bhuiyan, "A Lightweight CNN Model for Detecting Respiratory Diseases From Lung Auscultation Sounds Using EMD-CWT-Based Hybrid Scalogram," in *IEEE Journal of Biomedical and Health Informatics*, Vol. 25, no. 7, Jul. 2021.

[32]. M. Lech, M. Stolar, R. Bolia, M. Skinner, "Amplitude-Frequency Analysis of Emotional Speech Using Transfer Learning and Classification of Spectrogram Images," in *Advances in Science, Technology and Engineering Systems Journal*, Vol. 3, no. 4, pp. 363-371, 2018.

[33]. Y. LeCun, K. Kavukcuoglu and C. Farabet, "Convolutional networks and applications in vision," in Proc. *IEEE International Symposium on Circuits and Systems*, pp. 253-256, 2010.

[34]. W. Shi, J. Caballero, F. Husz´ar, J. Totz, A. P. Aitken, R. Bishop, D. Rueckert, and Z. Wang, "Real-time single image and video superresolution using an efficient sub-pixel convolutional neural network", in Proc. *IEEE conference on CVPR*, pp. 1874–1883, 2016.

[35]. C. Valentini-Botinhao et al., "Noisy speech database for training speech enhancement algorithms and tts models," Centre for Speech Technology Research (CSTR), School of Informatics, University of Edinburgh, 2017. [*Available online*]: https://datashare.is.ed.ac.uk/handle/10283/2791, accessed Dec. 2022.

[36]. C. Veaux, J. Yamagishi, and S. King, "The voice bank corpus: Design, collection and data analysis of a large regional accent speech database," in Proc. *International Conference Oriental COCOSDA. IEEE*, pp. 1–4, 2013.

[37]. J. Thiemann, N. Ito, and E. Vincent, "The diverse environments multichannel acoustic noise database: A database of multichannel environmental noise recordings," in JASA, Vol. 133, no. 5, pp. 3591–3591, 2013.

[38]. Yair Altman, "export_fig," [*Available online*]: (https://github.com/altmany/export_fig/releases/tag/v3.28), accessed on Dec. 23, 2022.

[39]. A. W. Rix, J. G. Beerends, M. P. Hollier and A. P. Hekstra, *"*Perceptual evaluation of speech quality (PESQ) – a new method for speech quality assessment of telephone networks and codecs," in IEEE, 2001.

[40]. C. H. Taal, R. C. Hendriks, R. Heusdens and J. Jensen, "A short-time objective intelligibility measure for time-frequency weighted noisy speech", in Proc. ICASSP, 2010.

[41]. MATLAB colrmaps. https://www.mathworks.com/help/matlab/ref/colormap.html

[42]. Choi, Hyeong-Seok, et al., "Phase-aware speech enhancement with deep complex u-net," in Proc. *International Conference on Learning Representations,* Vancouver CANADA, May 2018.

[43]. E. J. Nustede and J. Anemüller, "Single-Channel Speech Enhancement with Deep Complex U-Networks and Probabilistic Latent Space Models," in Proc. *IEEE International Conference on Acoustics, Speech and Signal Processing (ICASSP)*, Rhodes Island, Greece, pp. 1-5, 2023.